\documentclass{article}
\begin{document}
\author{Anatoly Konechny and  Albert Schwarz\\
 \\
Department of Mathematics, University of California \\
Davis, CA 95616 USA\\
konechny@math.ucdavis.edu,  schwarz@math.ucdavis.edu}
\title{\bf 1/4-BPS states on noncommutative tori}
\maketitle
\begin{abstract}
We give an explicit expression for classical 1/4-BPS fields in supersymmetric 
Yang-Mills theory on noncommutative tori. We use it to study quantum 1/4-BPS states. 
In particular we calculate the degeneracy of 1/4-BPS energy levels.
\end{abstract}
\large

\section{Introduction}

 Let $\theta$ be an antisymmetric bilinear form on a lattice ${\bf Z}^{d}$.
An algebra $T_{\theta}$ of functions on a noncommutative torus   
can be defined as 
an algebra with generators $U_{\bf k}$ labelled by lattice vectors 
${\bf k}\in {\bf Z}^{d}$ and satisfying the relations 
\begin{equation} \label{ntorus}
U_{{\bf k}_{1}}U_{{\bf k}_{2}}=exp(\pi i <{\bf k}_{1}\theta {\bf k}_{2}>)
U_{{\bf k}_{1} + {\bf k}_{2}} \, .
\end{equation}
Consider derivations $\delta_{i}$ of $T_{\theta}$ defined by the following relations 
$$
\delta_{j}U_{\bf k } = 2\pi i k_{j} \ U_{\bf k}, . 
$$
This derivations span an abelian Lie algebra which we denote by $L_{\theta}$. 
Given a homomorphism of a Lie algebra $L$ into $L_{\theta}$ we define a connection 
with respect to $L$ as follows. A connection $\nabla_{X}$ on a $T_{\theta}$ module $E$ 
is a set of linear operators $\nabla_{X} : E \to E$, $X\in L$ satisfying the Leibnitz rule
$$
\nabla_{X}(ae)=a\nabla_{X}e + (\delta_{X}a)e
$$
for any $a\in T_{\theta}$, $e\in E$. Here $\delta_{X}$ stands for the image of $X\in L$ in 
$L_{\theta}$ under the  homomorphism introduced above. We also assume that $\delta_{X}$ and 
$\nabla_{X}$ depend linearly on $X$.
Below we are interested in the case when $L$ is a ten-dimensional abelian Lie algebra endowed 
with a metric. In addition we fix a basis in $L$ consisting of vectors $X_{0}$, $X_{i}$, 
$i=1, \dots , d $, $X_{I}$, $I=d+1, \dots , 9$ such that $\delta_{X_{i}}=\delta_{i}$, 
$\delta_{I}=0$ and the metric  tensor $g_{\mu \nu}$ 
in this basis satisfies $g_{00}=-1$, $g_{i0}=g_{I0}=0$, $g_{iJ}=0$, $g_{IJ}=\delta_{IJ}$.
(Note that the index conventions in the present paper are different from the ones used in 
our preceding paper \cite{KS2}.)

If ${\bf Z}^{d}$ entering the definition of a 
noncommutative torus (\ref{ntorus}) is considered as a lattice $D$ in ${\bf R}^{d}$ one 
can associate to a given noncommutative torus $T_{\theta}$ two commutative 
tori $T^{d}$ and $\tilde T^{d}$.
 The torus $T^{d}$ 
is obtained  as a quotient $({\bf R}^{*})^{d}/D^{*}$ where $D^{*}$ is the dual
 lattice to $D$. In a more invariant way it can be described as a group of 
automorphisms of $T_{\theta}$ that corresponds to the Lie algebra
$L_{\theta}$. The metric tensor $g_{ij}$ determines a metric on $T^{d}$.
 The torus $\tilde T^{d}$ is defined as   ${\bf R}^{d}/D$. It can be
considered as a torus dual to $T^{d}$; it is equipped with a metric
specified by the matrix $g^{ij}$ that is inverse to
the matrix $g_{ij}$.   

It is shown in \cite{CDS} that toroidal compactification of matrix model (\cite{BFSS}, \cite{IKKT}) 
leads to Yang-Mills 
theory on a noncommutative torus. More precisely, the Minkowski action functional 
of compactified theory takes the form
$$
S= \frac{-V}{4g_{YM}^{2}}{\rm Tr} (F_{\mu\nu} + \phi_{\mu\nu}\cdot {\bf 1}) (F^{\mu\nu} + \phi^{\mu\nu}\cdot {\bf 1})  + \frac{iV}{2g_{YM}^{2}} {\rm Tr}\bar \psi \Gamma^{I}[\nabla_{I}, \psi ] 
\, .
$$
Here $F_{\mu\nu}$ is the curvature of connection 
$\nabla_{\mu}$ defined with 
respect to an abelian ten-dimensional Lie algebra $L$ on a 
projective module $E$ over $T_{\theta}$ ,  $\phi_{\mu\nu}$ is a constant 
antisymmetric tensor whose only non-vanishing components are $\phi_{ij}$, 
$i,j=1, \dots , d$, 
 $\psi$ is a ten dimensional Majorana-Weyl spinor taking values in the algebra of endomorphisms 
$End_{T_{\theta}}E$.  
 This action functional is invariant under the following supersymmetry transformations 
\begin{eqnarray} \label{susy}
&&\delta_{\epsilon, \tilde \epsilon} \nabla_{I} = \frac{i}{2}\bar \epsilon \Gamma_{I}\psi
\nonumber \\
&& \delta_{\epsilon, \tilde \epsilon} \psi = -\frac{1}{4} F_{IJ}\Gamma^{IJ}
\epsilon + \tilde \epsilon \cdot {\bf 1} \, .
\end{eqnarray}
  
It is clear from (\ref{susy}) that constant curvature connections on $T_{\theta}$ can be 
identified with 1/2-BPS fields (see \cite{CDS} for details). In  papers \cite{KS}, 
\cite{KS2} we studied constant curvature connections and quantum states that arise by quantization 
of these fields and their fluctuations. Using the constructions of \cite{KS} and 
\cite{KS2} one can obtain  not only the energies of  the 1/2-BPS states but also the energies of  
other quantum states, in particular,
1/4-BPS states. In the present paper we give a direct construction of 1/4-BPS classical 
fields and use it to study quantum 1/4-BPS states. We restrict ourselves to the case when the 
matrix $\theta_{ij}$ is irrational and, moreover, any linear combination of its entries with 
integer coefficients is irrational.
Most of our results do not depend on this restriction. However, it permits us to simplify 
drastically our considerations. Namely, using the above restriction we will be able to reduce 
the problems we consider to the case of a free module (the notion of Morita equivalence and 
the results of \cite{ASMorita}, \cite{ASRieffel}, \cite{KS} play a crucial role in this reduction).

Let us formulate our results first for the case $d=2$.  A projective 
module $E$
over a two-dimensional noncommutative torus $T_{\theta}$ can be 
characterized by means of two integers $p$ and $q$ obeying 
$p-q\theta >0$. Let us also assume that $p$ and $q$ are relatively prime. A 1/4-BPS state 
on $E$ is characterized by topological numbers $p$, $q$ and  integers 
$m_{i}, n^{j} \quad i,j=1,2$ that specify the eigenvalues of  operators 
${\cal P}_{i} = {\rm Tr}F_{ij}P^{j}$ and $p^{i}={\rm Tr}P^{i}$. (Here 
$P^{i}$  stands for the 
momentum canonically conjugated to $\nabla_{i}$ in the Hamiltonian formalism). 

We will obtain  the following expression for 1/4-BPS energy spectrum
\begin{eqnarray} \label{d=2spec}
&&E_{p,q,m_{i},n^{j}}
=\frac{( g_{YM})^{2}}{2VdimE}(n^{i} + \theta^{ik}m_{k})g_{ij}(n^{j} + \theta^{jl}m_{l}) 
+ \nonumber \\
&& + \frac{1}{2Vg_{YM}^{2}dimE}(\pi q + \phi_{12}dimE)^{2}
 + \frac{2\pi}{dimE} \sqrt{ K_{i}g^{ij}K_{j}}
\end{eqnarray}
where $K_{i} = pm_{i} - q_{ij}n^{j}$. 
The degeneracy of $E_{p,q,m_{i},n^{j}}$ is given by the number $c(K)$  where $K = g.c.d.(K_{i})$ and 
$c(K)$ is the coefficient at $x^{K}$ in the Taylor expansion of the function 
\begin{equation}\label{pf0}
Z(x) = 2^{8}\prod_{n}\left(\frac{1+x^{n}}{1-x^{n}}\right)^{8} \, . 
\end{equation}

For $d=3$ projective modules are labelled by an integer $p$ and an antisymmetric $3\times 3$ 
matrix $q_{ij}$ with integer entries. These numbers have to satisfy 
$dimE=p + \frac{1}{2}{\rm tr}\theta q > 0$.  
The 1/4-BPS spectrum reads
\begin{eqnarray} \label{d=3spec}
&&E=\frac{( g_{YM})^{2}}{2VdimE}(n^{i} + \theta^{ik}m_{k})g_{ij}(n^{j} + \theta^{jl}m_{l}) 
+ \nonumber \\
&& + \frac{V}{4dimEg_{YM}^{2}}(\pi q_{ij} + dimE\phi_{ij})g^{ik}g^{jl}(\pi q_{kl} + dimE\phi_{kl})
+ \nonumber \\
&& + \pi \sqrt{v_{i}g^{ij}v_{j}}
\end{eqnarray}
where $v_{i} = m_{i}dimE - q_{ij}(n^{j} + \theta^{jk}m_{k})$. 
The degeneracy of this eigenvalue is given by $c(K)$ obtained from (\ref{pf0}) for 
$K=g.c.d.(pm_{i} - q_{ij}n^{j}, m_{i}\frac{1}{2}\epsilon^{ijk}q_{jk})$. 
This expression agrees with the expression obtained in \cite{HV} for the commutative case.
We do not consider the case when $(p, q_{ij})$ are not relatively prime. The results for 
this case can be derived from our considerations combined with the results of \cite{DMVV}, \cite{DVV}. 
The eigenvalues (\ref{d=2spec}), (\ref{d=3spec}) coincide with the BPS bounds obtained in \cite{KS2} 
from supersymmetry algebra. We work in the Hamiltonian formalism on a three-dimensional noncommutative 
torus. Instead of that we could consider 1/4-BPS states using the Euclidean Lagrangian formalism on a 
four-dimensional noncommutative torus. In this setting the problem is related to the study of 
cohomology of the moduli space of noncommutative instantons. This means that one can obtain information 
about these cohomology groups from our calculations.

For $d>3$ we are able to analyze the degeneracy of 
1/4-BPS states only in modules admitting a constant curvature 
connection. (For $d=2,3$ every module has this property.) It was shown in \cite{KS}, \cite{AS} 
that such modules can be characterized by the property that corresponding K-theory class $\mu$ 
is a generalized quadratic exponent. We impose also an additional requirement that the module 
cannot be represented as a direct sum of equivalent modules (i.e. $\mu$ is not divisible by 
integer larger than 1 in the K-group). Such modules were called basic modules in \cite{KS}.
We calculate the spectrum of 1/4-BPS states and the corresponding multiplicities in basic modules.
The multiplicity again is given by a number $c(K)$ where the expression for $K$ in terms of 
topological numbers is given in section 3.


\section{Classical 1/4 BPS solutions}
We start with a brief discussion of the Hamiltonian formalism in the model 
at hand.
In \cite{KS} we discussed the quantization  of  Yang-Mills theory on a
noncommutative torus in the $\nabla_{0} = 0$ gauge.
Those considerations can be easily generalized to the supersymmetric case. 
Let us briefly describe the quantization procedure. The Minkowski
action functional  is defined on the configuration space 
$ConnE\times (\Pi End_{T_{\theta}}E)^{16}$
where $ConnE$ denotes the space of connections on $E$, $\Pi$ denotes the
parity reversion operator. To describe the Hamiltonian formulation we
first restrict ourselves to the space
${\cal M} = Conn' E\times (End_{T_{\theta}}E)^{9} \times (\Pi
End_{T_{\theta}}E)^{16}$
where $Conn'E$ stands for the space of connections satisfying
$\nabla_{0}=0$, the second factor
corresponds to a cotangent space to $Conn'E$. We denote coordinates on that cotangent 
space by $P^{I}$. 
Let ${\cal N}\subset {\cal M}$ be a subspace where the Gaussian constraint $[\nabla_{i},P^{i}]=0$
is satisfied.  
Then the  phase space of the theory is the quotient ${\cal P}={\cal N}/G$ where $G$
is the  group of spatial gauge transformations.

The presymplectic form (i.e. a degenerate closed 2-form)  $\omega$ on ${\cal M}$ is defined as
\begin{equation} \label{sform}
\omega = {\rm Tr} \delta P^{I}\wedge \delta \nabla_{I}
-i\frac{V}{2g_{YM}^{2}}{\rm Tr}\delta \bar \psi \Gamma^{0} \wedge \delta \psi \, .
\end{equation}
It descends to a symplectic form on the phase space $\cal P$.
The Hamiltonian of the theory reads 
\begin{eqnarray}\label{Hamiltonian}
&&H= {\rm Tr} \frac{g_{YM}^{2}R_{0}^{2}}{2V}
P^{I}g_{IJ}P^{J}  + \nonumber \\
&& +  {\rm Tr}\frac{V}{4g_{YM}^{2}}(F_{IJ} +
\phi_{IJ}\cdot {\bf 1})g^{IK} g^{JL}(F_{KL} +
\phi_{KL}\cdot {\bf 1}) +  \nonumber \\
&& + \mbox{fermionic terms} \, .
\end{eqnarray}

Let $\theta$ be an antisymmetric $d\times d$ matrix. We assume that any   linear combination 
of its entries  with integer coefficients is irrational.  
Consider   the corresponding  $d$-dimensional noncommutative torus  $T_{\theta}$.  
Let  $E$ be a projective module over it.  
In this paper we consider a particular class of modules  introduced in \cite{KS} which we named  
basic modules. A module $E$ is called basic if the algebra $End_{T_{\theta}}E$ is a
noncommutative torus $T_{\tilde \theta}$ and there is a constant curvature connection 
$\nabla_{i}$ on $E$ 
that satisfies the condition $[\nabla_{i}, \phi]=\tilde \delta_{i}\phi$ for every 
$\phi \in End_{T_{\theta}}E$ (here $\tilde \delta_{i}, i=1,\dots d$ is a basis of the 
algebra $L_{\tilde \theta}$ of derivations of $T_{\tilde \theta}$). 
The condition that $E$ is basic  is equivalent to the condition that  $T_{\theta}$ is 
completely Morita equivalent to the torus  $T_{\tilde 
\theta}=End_{T_{\theta}}E$ so that the module $\tilde E$ over 
$T_{\tilde \theta}$ corresponding to $E$ under this equivalence is a one dimensional free 
module. Therefore, provided components of $\theta$ satisfy the irrationality condition above, 
$\theta$ and  $\tilde \theta$ are related by an $SO(d,d|{\bf Z})$ transformation
\begin{equation} \label{Morita} 
\tilde \theta = (M\theta + N)(R\theta + S)^{-1} 
\end{equation}
where 
\begin{equation} \label{g}
g=\left(
\begin{array}{cc}
M&N\\
R&S
\end{array}
\right) \, \in SO(d,d|{\bf Z}) \, .
\end{equation}

As specified in the definition a basic module $E$ is equipped with a constant curvature 
connection. All modules over noncommutative tori admitting a constant curvature connection 
are classified in \cite{AS} (see also Appendix D of \cite{KS}) in terms of their 
representatives in the K-group $K_{0}(E)$. 
An element $\mu(E) \in K_{0}(T_{\theta})$ 
that corresponds to a  module admitting a constant curvature connection has the 
form of a generalized quadratic exponent.

If $T_{\theta}$ is a torus of dimension $d=2$ or $d=3$, 
 then an  element in $K_{0}(E)$ representing a module $E$ over $T_{\theta}$ 
is always a generalized quadratic exponent; it has the form 
$$
\mu(E) = p + \frac{1}{2}\alpha^{i}q_{ij}\alpha^{j}
$$ 
where $(p, q_{ij})$ are 
integers. The dimension of such a module is given by the formula 
$$
dimE=p + \frac{1}{2}tr(\theta q) \, .
$$
It follows from the results of \cite{ASMorita} that 
under an $SO(d,d|{\bf Z})$ transformation (\ref{Morita}), (\ref{g}) the numbers 
$(p, q_{ij})$ transform according to a spinor representation. 
The condition that $\mu(E)$ given above  corresponds to
 a basic module is  $gcd(p,q_{ij}) = 1$.

Let us fix a basic module $E$ over a $d$-dimensional torus $T_{\theta}$ 
and thus a generalized quadratic exponent $\mu = \mu(E)$. 
Denote by $Z_{\bf k}$, ${\bf k}\in \Gamma$ generators of the torus 
$T_{\tilde \theta} = End_{T_{\theta}}E$. They satisfy the  relations
\begin{equation} \label{Z}
Z_{{\bf k}}Z_{{\bf k}'}= e^{\pi i k_{i}\tilde \theta^{ij}k'_{j}} Z_{{\bf k}+{\bf k}' } \, .
\end{equation} 
 Using  the supersymmetry algebra of the model at hand (see \cite{KS2}) 
one can describe all classical  solutions preserving 
1/4 of all supersymmetries  which we  call 1/4 BPS fields. Our discussion of these solutions here 
essentially parallels the one in \cite{HV} for the commutative case. 
 
The  equations defining 1/4 BPS fields are of the following form    
\begin{eqnarray} \label{eq}
&& \psi = 0 \, , \nonumber \\
&& [\nabla_{i}, \nabla_{j}] = c_{ij} \cdot {\bf 1} \, , \quad  
[\nabla_{i}, X_{I}] = 0 \quad 
i,j = 1, \dots , d-1 \, , \quad [X_{I}, X_{J}]= 0 \nonumber \\
&& P^{d} = p^{d}\cdot {\bf 1} \, , \quad P_{i} = c_{i}\cdot {\bf 1} + 
Vg_{YM}^{-2}R_{d}^{-1}[\nabla_{d},\nabla_{i}] \, , \quad 
P_{I} = Vg_{YM}^{-2}R_{d}^{-1}[\nabla_{d}, X_{I}] \nonumber \\
&& \sum_{i=1}^{d-1}[\nabla_{i}, P^{i}] + \sum_{I=d+1}^{9}[X_{I}, P^{I}] = 0     
\end{eqnarray}
where $R_{d}^{-1}= \sqrt{g^{dd}}$, and $c_{ij}$, $c_{i}$, $p^{d}$  are constants.
More precisely, equations (\ref{eq}) define 1/4-BPS fields that are invariant under 
the supersymmetry transformations (\ref{susy}) with $\epsilon$, $\tilde \epsilon$ satisfying 
\begin{eqnarray}\label{str}
&&(\Gamma^{0} + \Gamma^{d})\epsilon = 0 \, , \nonumber \\
&& \tilde \epsilon = -\frac{1}{4}(c_{ij}\Gamma^{ij} + 2\Gamma^{0}\Gamma^{i}c_{i} + 
2\Gamma^{0}\Gamma_{d}p^{d})\epsilon \, .
\end{eqnarray}

In equations (\ref{eq}), (\ref{str}) we single out the $d$-th direction. In general a set of 
1/4-BPS fields preserved by the same subgroup of supersymmetries determines  a 
primitive lattice vector ${\bf m}\in {\bf Z}^{d}$. (The equations (\ref{eq}) are obtained
if one   changes a basis in the lattice 
${\bf Z}^{d}$ so that $\bf m$ is the $d$-th  basis vector.)  

The last equation in (\ref{eq}) is the Gauss constraint. 
The gauge equivalence classes of solutions to (\ref{eq}) define a subspace in the phase space of the 
theory - a moduli space of 1/4 BPS fields.  
The presymplectic form (\ref{sform}) gives rise to a symplectic form on the moduli space 
and thus 
one can perform a quantization of the resulting theory.  
As  any basic module (any module in the $d=2,3$ case)  
can be transformed by means of Morita equivalence to a free 
module we can restrict ourselves to consideration of a free module where the moduli space of 1/4 BPS fields 
can be most easily studied. Thus from now on we assume that $\tilde E$ is a rank 1 
free module over the torus $T_{\tilde \theta}=End_{T_{\theta}}E$.

It is easy to check that the following formulas define 1/4 BPS fields
\begin{eqnarray} \label{1/4'}
&&\nabla_{i}= \tilde \delta_{i} +  \sum_{k\in {\bf Z}} A_{i}(k) Z_{k{\bf m}}, 
\quad i=1, \dots , d-1 \nonumber \\
&& P^{j} = \tilde p^{j}\cdot {\bf 1} +  
\tilde Vg_{YM}^{-2}\tilde R_{d}^{-1}(\sum_{k\in {\bf Z}}ik A^{j}(k) Z_{k{\bf m}}) 
\quad i=1, \dots , d-1 \nonumber \\
&& P^{d} = \tilde p^{d}\cdot {\bf 1}
\, , \quad A_{d}=q_{d}\cdot {\bf{1}} \, , \quad \psi = 0 \nonumber \\
&&X_{I}=\sum_{k\in {\bf Z}-\{0\}} X_{I}(k) Z_{k{\bf m}} \nonumber \\
&& P^{I} = \tilde Vg_{YM}^{-2}\tilde R_{d}^{-1}(\sum_{k\in {\bf Z}}ik X^{I}(k) Z_{k{\bf m}}) 
\end{eqnarray}
where $\tilde \delta_{i}$ are derivations of $T_{\tilde \theta}$ acting according to 
the formula 
\begin{equation}\label{delta}
\tilde \delta_{j}Z_{\bf k} = 2\pi i k_{j} Z_{\bf k}\, .
\end{equation}
The last formula fixes a basis in the Lie algebra $L$, and 
$\tilde V$, $\tilde R_{d}$ refer to the metric tensor $\tilde g_{ij}$  in this basis.  
It is straightforward to check that the set of fields (\ref{1/4'}) also satisfies the Gauss constraint.
Moreover, any solution to (\ref{eq}) can be brought to the form (\ref{1/4'}) by means of 
gauge transformations. 
Let us sketch a proof of this fact.  
Any connection on $\tilde E$ has the form $\nabla_{\alpha}= \tilde \delta_{\alpha} + A_{\alpha}$, 
($\alpha=1, \dots , d$) where $A_{\alpha}$ are endomorphisms of $\tilde E$, which are just 
elements of the torus  $T_{\tilde \theta}$ acting on $\tilde E$ from the right (by a slight 
abuse of notation we denote  the generators of endomorphisms of $\tilde E$ by the same letters 
as generators of the torus $T_{\tilde \theta}$).  
First, because of the flatness condition $[\nabla_{i}, \nabla_{j}]=0$, $i,j=1, \dots , d-1$ 
we can bring the connection components $\nabla_{i}$ to the form 
$$
\nabla_{i}= \tilde \delta_{i} +  
\sum_{k\in {\bf Z}} A_{i}(k) Z_{k{\bf m}} \, .
$$
Due to the equation $ P_{i} = \tilde Vg_{YM}^{-2}\tilde R_{d}^{-1}[\nabla_{d},\nabla_{i}]$ 
the Gauss constraint $[\nabla_{i}, P^{i}] + [X_{I}, P^{I}] = 0$ now takes the form
$$
\nabla_{i}\nabla^{i}(A_{d}) + [X_{I}, [X^{I}, A_{d}]] = 0 \, .
$$
This  equation  implies that $A_{d}$ is of the form 
$$
A_{d}=\sum_{k\in {\bf Z}} A_{d}(k) Z_{k{\bf m}} \,.
$$
Finally, using gauge transformations generated by $ Z_{k{\bf m}}$ we can bring $A_{d}$ to the desired 
form $A_{d}=const\cdot {\bf{1}}$.

Note that the space of fields having the form (\ref{1/4'}) is still invariant under a  subgroup  of gauge 
transformations that consists of transformations defined by monomials $Z_{\bf k}$. 
We denote this group by $G^{mon}$. 
The symplectic form (\ref{sform}) being restricted to the subspace (\ref{1/4'})  reads 
\begin{eqnarray} \label{sform1/4}
&&\omega_{1/4} = \tilde Vg_{YM}^{-2}\tilde R_{d}^{-1}(\sum_{k\in {\bf Z}-\{0\}}ik 
( \sum_{j=1}^{d-1} \delta A^{j}(k)\wedge 
\delta A_{j}(-k) + \nonumber \\
&&+\sum_{I=d+1}^{9}\delta X_{I}(k)\wedge \delta X^{I}(-k))) + 
 \sum_{s=1}^{d}\delta \tilde p^{s}\wedge \delta q_{s} 
\end{eqnarray}
where $q_{s}$ stands for the zero mode component $A_{s}(0)$, $s=1, \dots , d$. 
The Hamiltonian has the form 
\begin{eqnarray} \label{H1/4}
&&H_{1/4}= \frac{g_{YM}^{2}}{2\tilde V}\sum_{s,r=1}^{d}\tilde p^{s}\tilde g_{sr}\tilde p^{r} 
+ \frac{\tilde V}{4g_{YM}^{2}}  \phi_{\mu\nu}\phi^{\mu\nu} + \nonumber \\
&& + \tilde Vg_{YM}^{-2}\tilde R_{d}^{-2}(\sum_{k\in {\bf Z}-\{0\}} k^{2} (\sum_{i=1}^{d-1}A_{i}(k)A^{i}(-k) + 
\sum_{I=d+1}^{9}X_{I}(k)X^{I}(-k) )) \, .
\end{eqnarray}

The space of fields  of the form (\ref{1/4'}) can be extended by adding fermionic degrees of freedom to 
a minimal supermanifold ${\cal B}_{1/4}$ invariant under all supersymmetry transformations. This is achieved by 
adding spinor fields $\psi = \sum_{k\in Z} \psi (k) Z_{k{\bf m}}$. The corresponding   additional term 
to the Hamiltonian (\ref{H1/4}) reads 
\begin{equation} \label{Hferm}
H_{1/4}^{ferm} = \frac{V}{2g_{YM}^{2}} \sum_{k\in {\bf Z}} ik  \psi(k)^{t} \Gamma^{d}\psi (k) \, .
\end{equation} 

To describe 1/4-BPS states we should quantize the systems described by the Hamiltonian 
$H_{1/4} + H_{1/4}^{ferm}$ on a symplectic manifold ${\cal B}_{1/4} / G^{mon}$. 
(We can obtain the 
symplectic form on this manifold restricting the form (\ref{sform}) or adding fermionic terms 
to (\ref{sform1/4}).)
The combined system (\ref{H1/4}), (\ref{Hferm}) describes  a free motion  for 
the zero modes degrees of freedom $q_{s}$, $\psi(0)$,  
and an infinite system of supersymmetric harmonic oscillators with frequences 
$\omega (k) = \|k\| = \sqrt{k\tilde g^{dd}k}$.   
This system is a direct  analogue of the chiral sigma model on ${\bf R}^{6}\times  T^{2}$ 
considered in \cite{HV}, \cite{HMS}.  
More precisely, we can introduce periodic functions 
$$
A_{i}(\phi)=\sum_{k\in {\bf Z}} A_{i}(k)e^{ik\phi} \quad 
X_{I}(\phi)=\sum_{k\in {\bf Z}} X_{I}(k)e^{ik\phi} \, .
$$  
and express the Hamiltonian and symplectic form in terms of these functions:
\begin{eqnarray} \label{Hsigma}
&& H= \frac{g_{YM}^{2}}{2\tilde V}\sum_{s,r=1}^{d}\tilde p^{s}\tilde g_{sr}\tilde p^{r} 
+ \frac{\tilde V}{4g_{YM}^{2}}  \phi_{\mu\nu}\phi^{\mu\nu} + \nonumber \\
&& + \tilde Vg_{YM}^{-2}\tilde R_{d}^{-2} \int_{0}^{2\pi} d\phi \left( 
\sum_{i=1}^{d-1}\frac{d A^{i}}{d\phi} \frac{d A_{i}}{d\phi} + \sum_{I=d+1}^{9} 
\frac{d X^{I}}{d\phi} \frac{d X_{I}}{d\phi} \right) \, ,
\end{eqnarray} 
\begin{eqnarray} \label{omegasig}
&&\omega = 
\tilde Vg_{YM}^{-2}\tilde R_{d}^{-1}\int_{0}^{2\pi}d\phi 
\left( \sum_{j=1}^{d-1}\frac{d (\delta A^{j})}{d\phi} \wedge \delta' 
A_{j}   
+ \sum_{I=d+1}^{9} \frac{d (\delta X^{I})}{d\phi} \wedge \delta' X_{I}  
\right) + \nonumber \\ 
 && + \sum_{s=1}^{d}\delta \tilde p^{s}\wedge \delta q_{s} \, . 
\end{eqnarray}
We obtained the Hamiltonian and symplectic form of the standard chiral sigma-model.
(We write down only the bosonic part of the model. Its supersymmetrization is straightforward.)
However the phase space is not standard. We should factorize with respect to the 
action of group $G^{mon}$. In other words we identify $A_{j}(\phi)$ with 
$A_{j}(\phi + 2\pi  \tilde \theta ^{ds}n_{s} ) + 2\pi i n_{j}$ and 
$X_{I}(\phi)$ with $X_{I}(\phi + 2\pi  \tilde \theta ^{ds}n_{s} )$
for any integer valued vector $(n_{s}) \in {\bf Z}^{d}$ that specifies an element of $G^{mon}$. 
For $\theta = 0$ this means that we consider the classical sigma-model on 
$T^{d-1}\times {\bf R}^{9-d}$; we will use this terminology also in the case $\theta \ne 0$ 
although one should emphasize that our sigma-model is not completely standard.


\section{Quantization}

For now let us concentrate on the bosonic part (\ref{H1/4}) of the model. 
Upon the Hamiltonian quantization of  the model (\ref{H1/4}), (\ref{sform1/4}) 
we get a Hilbert space spanned 
by the wave functions 
\begin{equation}\label{wfunction}
\Psi_{\tilde p_{i};N_{1}(k),\dots , N_{8}(k)} = exp(i\sum_{s=1}^{d}\tilde p^{s}q_{s})
\prod_{k\in {\bf N}} 
(a^{\dagger}_{1}(k))^{N_{1}(k)}\cdot \dots \cdot (a^{\dagger}_{8}(k))^{N_{d-1}(k)}|0\rangle 
\end{equation}
where $a^{\dagger}_{i}(k)$ ($k$ is a natural number) are oscillators creation operators, $N_{i}(k)$ 
are the corresponding occupation numbers, and $|0\rangle $ stands for the oscillators ground state.

Using (\ref{Z}) and (\ref{delta}) one can calculate the action of the group $G^{mon}$ 
on the wave function (\ref{wfunction}). 
An element $Z_{\bf n}\in G^{mon}$ acts on  (\ref{wfunction}) 
by multiplication by the exponential factor 
$$
exp\left( 2\pi i\sum_{s=1}^{d}n_{s}(-\tilde p^{s} + \tilde \theta^{sd} \sum_{k\in {\bf N}} 
\sum_{l=1}^{8}N_{l}(k)k) \right) \, .
$$
Hence, the invariance of state vectors under the gauge transformations generated by the group 
 $G^{mon}$ leads to the quantization condition of zero modes $\tilde p^{s}$:
\begin{equation}\label{tildep}
\tilde p^{s} = \tilde n^{s} + \tilde \theta^{sd} \sum_{k\in {\bf N}} 
\sum_{l=1}^{8}N_{l}(k)k
\end{equation}
where $\tilde n^{s}$ are integers. 
Quantization of the fermionic part (\ref{Hferm}) is straightforward. Note that the zero modes 
$\psi (0)$ are not dynamical. They only  influence the degeneracy of states.

The energy spectrum of the system reads 
\begin{equation}\label{E}
 E_{\tilde n_{s}, K} =  \frac{g_{YM}^{2}}{2V}
(\tilde n^{s} + \tilde \theta^{sd}K)g_{sr}( \tilde n^{r} + 
\tilde \theta^{rd}K) +  \frac{V}{4g_{YM}^{2}}  \phi_{\mu\nu}\phi^{\mu\nu}
 + \| K\| 
\end{equation}
where 
\begin{equation} \label{K}
K= \sum_{k\in {\bf N}} \sum_{l=1}^{8}N_{l}(k)k + \sum_{k\in {\bf N}} \sum_{i=1}^{8} {\cal V}_{i} (k) k \, .
\end{equation}
In the last formula ${\cal V}_{i} (k) = 0,1 $ stand for fermionic  occupation numbers. 
For fixed numbers $\tilde n^{s}$ and $K$ the degeneracy of the energy 
eigenvalue (\ref{E}) is defined  by the number of representations of $K$ in the form (\ref{K}) (the 
number of partitions).
 More explicitly,  the degeneracy of the eigenvalue $E_{\tilde n_{i}, K}$ is given by 
the coefficient at the $K$-th power of $x$ in the partition function
\begin{equation} \label{pf}
Z(x) = 2^{8}\prod_{n}\left(\frac{1+x^{n}}{1-x^{n}}\right)^{8} \, .
\end{equation}

From  general arguments (see \cite{HofVer2}, \cite{KS2}) we know that the eigenvalues of 
operators  $p^{i} = {\rm Tr}P^{i}$ and ${\cal P}_{i} = {\rm Tr}F_{ij}P^{j}$ obey the following 
quantization conditions
$$
p^{i} = n^{i} + \theta^{ij}m_{j}
$$
$$
{\cal P}_{j} = 2\pi m_{j}
$$
where $n^{i}$, $m_{j}$ ($i,j=1, \dots d$) are integers.
These eigenvalues are defined for a given  module $E$ and connection $\nabla_{\alpha}$ on it. 
In transition from $E$ to  a module $\tilde E$ over Morita equivalent torus $T_{\tilde \theta}$
connections and endomorphisms on $E$ are mapped to connections and endomorphisms on $\tilde E$. 
Thus,   the integers   $n^{i}$, $m_{j}$ defining the eigenvalues of operators  $p^{i}$ and 
${\cal P}_{i}$ on $E$ transform  to new integers $\tilde n^{i}$ and $\tilde  m_{j}$ related to $\tilde E$. 
One can prove (see \cite{HofVer2}, \cite{KS}, \cite{KS2} ) that  the set of numbers $(-n^{i}, m_{j})$ 
transform according to a vector representation of the group 
$SO(d,d|{\bf Z})$.
For the case at hand when $E$ is some basic module over $T_{\theta}$ and $\tilde E$ is a free module 
over $T_{\tilde \theta}=End_{T_{\theta}}E$ the eigenvalues  $\tilde n^{i}$ are 
given in  (\ref{tildep}), and for $\tilde {\cal P}_{j}$ a straightforward calculation yields 
\begin{equation}\label{tildem}
\tilde {\cal P}_{j}=2\pi \tilde m_{j};\quad 
\tilde m_{j} = \delta_{jd} K  
\end{equation}
where $K$ is given by (\ref{K}).
 The integers $(\tilde n^{i}, \tilde m_{j})$ related to the free module 
$\tilde E$ can be expressed via the integers $n^{i}$ and  $m_{j}$ corresponding to $E$ 
as follows
\begin{equation} \label{n}
\tilde n^{i} = M^{i}_{j}n^{j} - N^{ij}m_{j}
\end{equation}
\begin{equation}\label{m}
\tilde m_{i} = S_{i}^{j}m_{j} - R_{ij}n^{j} \, .
\end{equation}
Conversely, the values of $n^{i}$ and  $m_{j}$  are related to 
 $(\tilde n^{i}, \tilde m_{j})$ by means of   the inverse matrix to  (\ref{g}): 
$$
g^{-1}=\left(
\begin{array}{cc}
S^{t}&N^{t}\\
R^{t}&M^{t}
\end{array} \right) \, .
$$
In transition from $\tilde E$ to $E$ one should also take into account the change of metric tensors 
$$
\tilde g = AgA^{t} \, , \quad A = R\theta + S \, , 
$$
transformation of the background field $\phi_{\mu \nu}$ 
$$
\tilde \phi = A\phi A^{t} + \pi RA^{t} \, ,
$$
and change of the coupling constant 
$$
\tilde g_{YM} = |det A|^{1/2}g_{YM}^{2}
$$
(see \cite{MorZum}, \cite{BrMorZum}, \cite{KS},  \cite{KS2} for the details).
The energy of BPS states can now be written in terms of the topological numbers specified 
by the matrix (\ref{g}) and quantum numbers  $n^{i}$,  $m_{j}$ as follows
\begin{eqnarray} \label{BPSspec}
&&E=\frac{( g_{YM})^{2}}{2VdimE}(n^{i} + \theta^{ik}m_{k})g_{ij}(n^{j} + \theta^{jl}m_{l}) 
+ \nonumber \\
&& + \frac{VdimE}{4g_{YM}^{2}}(\pi (A^{-1}R)_{ij} + \phi_{ij})g^{ik}g^{jl}(\pi (A^{-1}R)_{kl} + \phi_{kl})
+ \nonumber \\
&& + \pi \sqrt{v_{i}g^{ij}v_{j}}
\end{eqnarray}
where $v_{i}=A^{-1}_{ij}(S_{j}^{k}m_{k} - R_{jk}n^{k})$, $A=R\theta + S$ and $R$ and $S$ 
are blocks of the matrix (\ref{g}).  
Note that the expression (\ref{BPSspec}) does not refer to a particular choice of basis 
in the lattice ${\bf Z}^{d}$ which was used earlier for the sake of convenience. 
We skipped many technical details here as the calculation essentially parallels the one made in 
\cite{KS} for the case $d=2$. We also omitted the possible topological terms in formulas 
(\ref{E}), (\ref{BPSspec}). One can easily restore them. 
For the cases $d=2,3$ one can  rewrite the expression (\ref{BPSspec}) for BPS energy 
spectrum in terms of the topological numbers $(p, q_{ij})$ (formulas 
(\ref{d=2spec}) and (\ref{d=3spec}) in the Introduction to this paper). The 
answer coincides 
with the one given in \cite{KS2} and agrees with \cite{HofVer1}, \cite{HofVer2}.

Using (\ref{pf}), (\ref{tildem}), and (\ref{m}) one can also express the degeneracy 
of BPS states in terms of the numbers related to $E$. It is easy to see that the expression 
$g.c.d.(x_{i})$ where $x=(x_{1}, \dots , x_{d}) \in {\bf Z}^{d}$ is invariant under $SL(d,{\bf Z})$ 
transformations. From (\ref{tildem}) we know the answer for the case when the vector $\tilde m_{i}$ is 
directed along the $d$-th axis. Using an $SL(d,{\bf Z})$ transformation and invariance of 
$g.c.d.(\tilde m_{i})$ we obtain that in the general case the 
 degeneracy of BPS states is  specified by the number of partitions of the integer 
$$
K = g.c.d. (S_{i}^{j}m_{j} - R_{ij}n^{j}) \, .
$$  
For the cases $d=2$ and $d=3$ it is easy to find an explicit formula for  a matrix (\ref{g}) 
in terms of the topological numbers  $(p, q_{ij})$. For the two-dimensional case 
one can take 
$$
g=\left(
\begin{array}{cc}
p' \cdot {\bf 1}& q'_{ij}\\
 q_{ij}&  p \cdot {\bf 1}    
\end{array} \right)
$$  
where $\bf 1$ is the $2\times 2$ identity matrix, $q_{ij}'$ is an antisymmetric  $2\times 2$ matrix 
with integer entries and     $p'$ is an integer such that $q_{12}q_{12}' + pp' = 1$. 
Therefore, for $d=2$ the degeneracy of 1/4 BPS states is determined by 
$K = g.c.d. (pm_{i} - q_{ij}n^{j})$.

For $d=3$ one can derive the following expression for $\tilde m_{i}$ in terms of the topological 
numbers $p$, $q_{ij}$ and integers $m_{i}$, $n^{j}$
$$
\tilde m_{i} = S_{i}^{j}m_{j} - R_{ij}n^{j} = (p\delta_{j}^{i} + (1-p)\bar q_{i} q^{j}) m_{j} - q_{ij}n^{j} 
$$
where $q^{j} = \frac{1}{2}\epsilon^{jkl}q_{kl} (g.c.d.(q_{ij}))^{-1}$ and the integers $\bar q_{i}$ 
satisfy $\sum_{i=1}^{3}q^{i}\bar q_{i} = 1$. For $K=g.c.d. (\tilde m_{i})$  one 
 can write  a more compact formula $K = g.c.d.(pm_{i} - q_{ij}n^{j}, m_{i}\frac{1}{2}\epsilon^{ijk}q_{jk})$. 
Here we used the fact that the numbers $(p, q_{ij})$ are relatively prime. Using the same fact 
one can verify that  
$$
K = g.c.d.(pm_{i} - q_{ij}n^{j}, m_{i}\frac{1}{2}\epsilon^{ijk}q_{jk}) = 
g.c.d. (pm_{i} - q_{ij}n^{j}, m_{i}\frac{1}{2}\epsilon^{ijk}q_{jk}, m_{i}n^{i}) \, . 
$$
The last expression agrees with the formula  given in \cite{HV} for degeneracies of 1/4-BPS states on 
a three-dimensional commutative torus.

Let us discuss now how one can generalize our considerations to the case of 
arbitrary modules 
admitting a constant curvature connection. In this case one can reduce the problem to the consideration 
of BPS states in a free module having rank $N>1$. Again, we can describe 
1/4-BPS fields 
and reduce our problem to the quantization of an analogue of the orbifold chiral sigma model on 
$({\bf R}^{6}\times T^{2})^{N}/S_{N}$. More precisely, the general formula for 1/4-BPS fields 
in a free module of rank $N>1$ can be written in the same way as in the case $N=1$ 
(formula (\ref{1/4'})). The only difference is that in the general case $A_{i}(k)$, $X_{I}(k)$  
and the zero modes $q_{s}\equiv A_{s}(0)$, 
$\tilde p^{s}$ are $N\times N$ matrices satisfying the conditions
\begin{eqnarray}\label{cc}
&&\sum_{k + k' = n} [A_{i}(k), A_{j}(k')] = 0\nonumber \\
&&\sum_{k + k' = n} [A_{i}(k), X_{I}(k')] = 0 \nonumber \\
&&\sum_{k + k' = n} [X_{I}(k), X_{J}(k')] = 0 
\end{eqnarray}
for any $n\in {\bf Z}$ and any $i,j =1, \dots, d$, $I,J=d+1, \dots 9$. 
Introducing generating functions 
$$
A_{i}(\phi)=\sum_{k\in {\bf Z}} A_{i}(k)e^{ik\phi} \quad 
X_{I}(\phi)=\sum_{k\in {\bf Z}} X_{I}(k)e^{ik\phi} \, .
$$
we can express these conditions as commutation relations 
\begin{equation} \label{cr}
[A_{i}(\phi), A_{j}(\phi)] = 0 \, , \quad [A_{i}(\phi), X_{I}(\phi)]=0 \, , \quad
[X_{I}(\phi), X_{J}(\phi)]=0 \, .
\end{equation}
One can use the remaining gauge invariance to simplify further the study of BPS 
states. Using the commutation relations (\ref{cr}) one can prove that there exists such a 
matrix-valued function $u(\phi )$ satisfying the conditions that the matrices 
$u^{-1}(\phi)A_{i}(\phi )u(\phi )$, $u^{-1}(\phi )X_{I}(\phi )u(\phi)$  are diagonal and 
$u(0)=1$. In the generic case (more precisely in the case when for every $\phi$ at least one of 
the matrices $A_{i}(\phi )$, $X_{I}(\phi )$ has distinct eigenvalues) there exists a unique 
continuous function $u(\phi )$ obeying these conditions and $u(2\pi )$ is a permutation 
matrix. Let us stress that diagonalized matrix-valued functions $A_{i}(\phi )$, $X_{I}(\phi )$
are not periodic in general.

The above diagonalization permits us to reduce the study 
of 1/4-BPS states in an arbitrary free module to the study of an orbifold sigma 
model with the target space  $({\bf R}^{9-d}\times  T^{d-1})^{N}/S_{N}$.
  A model of this kind was analyzed  in \cite{DMVV}, \cite{DVV}. It was shown
there that  when $N$ tends to infinity such a model is related to string
theory in the light cone gauge. As it was emphasized at the end of Section 2
our sigma-model  is not completely standard, however 
the considerations of the papers we mentioned  can be applied to our
situation. We can conclude that in the situation when BPS-fields are
dominant the SYM theory on noncommutative torus is related to
string theory/M-theory. Looking more closely at this relation we find
an  agreement with the physical interpretation of compactifications
on noncommutative tori in terms of toroidal compactifications of M-theory
with non-vanishing expectation value of the three-form (see \cite{CDS}).  
In particular, the relation between M-theory/String spectra and SYM
spectra (for example see \cite{OP}) agrees with our formulas.

\begin{center} {\bf Acknowledgments} \end{center}
Most of this work was accomplished during the authors stay at the Mittag-Leffler Institute 
the warm hospitality of which is greatly acknowledged.  
We are indebted to R. Dijkgraaf, M. Douglas, C. Hofman, N. Nekrasov and M. Rieffel for
useful 
discussions.


\begin{thebibliography}{99}
\bibitem{BFSS} T.~Banks, W.~Fischler, S.~Shenker and L.~Susskind, {\it M-Theory as a
Matrix Model: A Conjecture}, Phys. Rev. {\bf D55} (1997) 5112, hep-th/9610043.
\bibitem{IKKT} N.~Ishibashi, H.~Kawai, I.~Kitazawa, and A.~Tsuchiya, 
{\it A large-N reduced model as superstring}, Nucl. Phys. B492 (1997) 
pp. 467-491; hep-th/9612115. 
\bibitem{CDS} A.~Connes, M.~Douglas and A.~Schwarz, {\it Noncommutative geometry and Matrix
theory: compactification on tori}, JHEP 02(1998)003; hep-th/9711162.
\bibitem{ASMorita} A.~Schwarz, {\it Morita equivalence and duality}, Nucl. Phys. B534 (1998) 
pp. 720-738; hep-th/9805034.
\bibitem{ASRieffel} M.~Rieffel and A.~Schwarz, {\it Morita equivalence of
multidimensional noncommutative tori}, to appear in the Int. 
Jr. of Mathematics; q-alg/9803057.
\bibitem{KS} A.~Konechny and A.~Schwarz, {\it BPS states on noncommutative
tori and duality}, Nucl. Phys. B550 (1999) 561-584; hep-th/9811159.
\bibitem{KS2} A.~Konechny and A.~Schwarz, {\it Supersymmetry algebra and BPS states of 
super Yang-Mills theories on noncommutative tori},  Phys. Lett. B453 (1999) 23-29; hep-th/9901077.
\bibitem{AS} A.~Astashkevich and A.~Schwarz, {\it Projective modules over noncommutative 
tori: classification of modules with constant curvature connection}, 
math.QA/9904139.
\bibitem{HofVer1} C.~Hofman and E.~Verlinde, {\it U-Duality of Born-Infeld on the Noncommutative Two-Torus}, JHEP 9812 (1998) 010; hep-th/9810116.
\bibitem{HofVer2} C.~Hofman and E.~Verlinde, {\it Gauge bundles and Born-Infeld on the
noncommutative torus}, Nucl. Phys. B547 (1999) 157-178; hep-th/9810219.
\bibitem{MorZum} B.~Morariu and B.~Zumino {\it Super Yang-Mills on the Noncomutative Torus}
to appear in the Arnowitt Festschrift Volume ``Relativity, Particle Physics, and
     Cosmology'', Texas A\& M University, April 1998; hep-th/9807198.
\bibitem{BrMorZum} D. Brace, B. Morariu and  B. Zumino {\it Dualities of 
the Matrix Model from T-Duality of the Type II String}, 
Nucl. Phys. B545 (1999) 192-216; hep-th/9810099.
\bibitem{HV} F. Hacquebord and H. Verlinde, {\it Duality Symmetry of N=4 Yang-Mills Theory on $T^{3}$}, 
Nucl. Phys. B508 (1997) 609-622; hep-th/9707179.
\bibitem{HMS} J. Harvey, G. Moore and A. Strominger, {\it Reducing S-duality to T-duality}, Phys. 
Rev. D52 (1995) 7161; hep-th/9501022.
\bibitem{DMVV} R. Dijkgraaf, G.~Moore, E.~Verlinde and H.~Verlinde, {\it Elliptic genera of symmetric 
products and second quantized strings}, Commun. Math. Phys. 185 (1997) p. 197; hep-th/9608096. 
\bibitem{DVV}  R. Dijkgraaf,  E.~Verlinde and H.~Verlinde, {\it Matrix 
string theory}, Nucl. Phys. B500 (1997) 43-61; hep-th/9703030.
\bibitem{OP} N.A. Obers and B. Pioline, {\it U-duality and M-theory}, 
to appear in Phys. Rept.; hep-th/9809039. 

\end{thebibliography}
\end{document}